\documentclass[aps,prl,superscriptaddress,twocolumn,floatfix]{revtex4-2}
\usepackage[abs]{overpic}
\usepackage{graphicx,graphics}
\usepackage{dcolumn}
\usepackage{amsmath,amssymb,amsfonts}
\usepackage{wasysym}
\usepackage{latexsym,verbatim}
\usepackage{bm}
\usepackage[dvipsnames]{xcolor}
\usepackage[framemethod=default]{mdframed}
\usepackage[breaklinks=true,colorlinks,citecolor=blue,linkcolor=blue,urlcolor=blue]{hyperref}
\usepackage[makeroom]{cancel}
\usepackage{fancybox}
\usepackage{ulem}
\usepackage{tikz}
\usetikzlibrary{decorations.pathreplacing}

\newcommand{\be}{\begin{eqnarray}}
\newcommand{\ee}{\end{eqnarray}}
\newcommand{\nn}{\nonumber\\}

\newcommand{\sgn}{\text{sgn}}

\begin{document}

\title{%Lattice-Commensurate Coupled-Wire Constructions of Pfaffian and Anti-Pfaffian States
%or 
Lattice Commensurability and Majorana sectors of Pfaffian–Anti-Pfaffian interfaces}

%Lattice Commensurability in Coupled-Wire Realizations of Pfaffian and Anti-Pfaffian States

\author{Feng Liu}
%\email{liufengz@ust.hk}
\author{Xue-Yang Song}
%\email{songxy@ust.hk}
\affiliation{Department of Physics, The Hong Kong University of Science and Technology, Clear Water Bay, Hong Kong SAR, China}

\begin{abstract}
We study Pfaffian and anti-Pfaffian non-Abelian fractional quantum anomalous Hall states in half-filled moir\'{e} lattice systems. 
Owing to crystal-momentum conservation modulo a reciprocal lattice vector, these lattice systems admit coupled-wire constructions of the two phases without requiring spatially varying magnetic-field-induced momentum shifts.
For the resulting Pfaffian--anti-Pfaffian interface, a Umklapp process can gap the counter-propagating charge sector, while the remaining four co-propagating Majorana modes naturally group into two sectors organized by translation symmetries.
Our results show that lattice commensurability qualitatively changes the coupled-wire accessibility of non-Abelian states while providing a microscopic momentum structure that directly constrains their edge and interface dynamics.
\end{abstract}

\maketitle

{\it Introduction}---Moir\'{e} materials provide a versatile platform for interaction-driven phases because their narrow bands strongly enhance correlation effects~\cite{bistritzer2011moire,nuckolls2024microscopic,bernevig2025fractional}, leading to correlated insulators~\cite{cao2018correlated,xu2020correlated,huang2021correlated}, superconductors~\cite{cao2018unconventional,balents2020superconductivity,xia2026bandwidth}, and fractional quantum anomalous Hall (FQAH) states~\cite{xu2023observation,lu2024fractional,cai2023signatures}.
More recently, two dimensional arrays of one-dimensional Luttinger liquids have been experimentally realized in moir\'{e} superlattices~\cite{wang2022one,li2024imaging,kawakami2026one}, providing a physical setting closely connected to coupled-wire models. 
Since coupled-wire constructions use interacting one dimensional modes as building blocks for two dimensional correlated and topological phases~\cite{vishwanath2001two,mukhopadhyay2001sliding,kane2002fractional,teo2014luttinger,shavit2024quantum,hu2024twisted,wu2024possible,wu2019coupled}, these developments motivate the use of coupled-wire approaches to explore topological states in moir\'{e} systems~\cite{hsu2023general}.

The coupled-wire construction provides a direct microscopic representation of the low energy edge operators in terms of the underlying electron degrees of freedom.
In particular, it keeps track of the microscopic momentum assignments of the edge modes~\cite{kane2002fractional,teo2014luttinger,neupert2014wire}. 
This information becomes especially relevant at interfaces, where the allowed couplings between edge modes are constrained by momentum selection rules~\cite{kane1994randomness,wan2016striped,barkeshli2015particle,lian2018theory,wang2018topological,mross2018theory,wang2013edge}. 
A prominent example is the interface between Pfaffian (Pf) and anti-Pfaffian (aPf) phases, where the momentum organization of the neutral Majorana modes constrains their possible hybridization and can qualitatively modify the resulting low energy phase structure and thermal Hall conductance~\cite{lian2018theory,wang2018topological,mross2018theory,zhu2020topological}.

These momentum sensitive interface phenomena motivate a microscopic understanding of how the underlying non-Abelian states can be realized in a lattice system. 
For a continuum quantum Hall system, the tunneling processes required in the coupled-wire construction of the fermionic Moore–Read state cannot simultaneously satisfy the momentum conservation conditions in a uniform magnetic field~\cite{teo2014luttinger,fuji2019quantum}. 
A spatially modulated magnetic field is therefore introduced to render all the required tunneling processes non-oscillatory~\cite{teo2014luttinger}.
In a lattice system, by contrast, crystal momentum is conserved only modulo a reciprocal lattice vector~\cite{hsu2023general,shavit2024quantum}, which enlarges the set of tunneling processes and relaxes the continuum momentum conserving conditions. 
This difference raises the possibility of realizing the Pfaffian state in a zero-field moir\'{e} system without introducing the spatially modulated magnetic field.

In this work, we exploit the lattice momentum structure of half-filled coupled-wire systems to construct enlarged wire unit cells that realize the Pfaffian and anti-Pfaffian phases. 
The enlarged construction admits a complete set of tunneling operators compatible with crystal momentum conservation, thereby reproducing the corresponding non-Abelian edge structures without requiring the spatially modulated magnetic field. 
We then use the microscopic edge information encoded in these constructions to analyze a Pf--aPf domain wall. 
Crystal momentum conservation allows an Umklapp process to fully gap the counter-propagating charge sector~\cite{wan2016striped,barkeshli2015particle}, while the remaining four co-propagating Majorana modes separate into two momentum sectors. 
This $2+2$ structure provides a microscopic basis for discussing the flavor symmetries that play an important role in effective theories of Pf--aPf domain walls~\cite{lian2018theory,wang2018topological,mross2018theory}.
More broadly, our results show that lattice momentum conservation can support coupled-wire constructions that are inaccessible under the continuum setup. 
By retaining the microscopic operator and momentum structure of the boundary modes, such construction provides the framework for studying momentum sensitive edge and interface phenomena in non-Abelian FQAH phases and may be extended to other non-Abelian states~\cite{fuji2017non,tam2020coupled,sirota2019paired}.

{\it Coupled-wire construction}---We begin by briefly reviewing the coupled-wire construction of the Pfaffian state in continuum quantum Hall systems~\cite{teo2014luttinger}. 
In this approach, the two dimensional system is represented as an array of parallel one dimensional quantum wires labeled by $j$, each described at low energies by a Luttinger liquid. 
The right- and left-moving electron operators are bosonized as~\cite{kane2002fractional,teo2014luttinger,fuji2019quantum}
\be
\psi_{R/L,j}(x)&=&\frac{\kappa_j}{\sqrt{2\pi \alpha}}\exp(i(k_{Fj}^{R/L}x+\varphi_j(x)\pm\theta_j(x)))
\ee
where $\alpha$ is a short distance cutoff, $\kappa_j$ is a Klein factor. 
The momentum assignments distinguish the continuum and lattice settings. 
In the continuum quantum Hall system, the background magnetic field shifts the Fermi points of neighboring wires according to $k_{Fj}^{R/L}=bj\pm k_F$, with $b$ encoding the momentum offset induced by the background magnetic field.
At filling $\nu$, the density is related to these momentum scales by $2k_F=\nu b$~\cite{kane2002fractional}.

Moir\'{e} systems provide a physical setting for lattice coupled-wire descriptions, as suggested by theoretical predictions of one-dimensional conducting channels along stacking domain walls~\cite{hsu2023general,chou2019superconductor} and experimental observations of Luttinger-liquid behavior~\cite{wang2022one}.
In the lattice setting, crystal momentum along the wires is conserved modulo the longitudinal reciprocal lattice vector $G=2\pi/\lambda$, where $\lambda$ is the period along the effective wires~\footnote{In the moir\'{e} quantum-network geometry~\cite{hsu2023general}, the wires run along lattice directions connecting neighboring AA regions, so their longitudinal period coincides with the moir\'{e} superlattice constant $\lambda$. 
Accordingly, $G$ corresponds to the projection of an appropriate two-dimensional reciprocal lattice vector onto the wire direction, rather than its magnitude.}.
We take the interwire momentum shift to vanish modulo $G$.
At filling $\nu$, the electron density per moir\'{e} unit cell fixes the Fermi momentum through $2k_F=\nu2\pi/\lambda$ ~\cite{giamarchi2003quantum,shavit2024quantum,hsu2023general}.
For the continuum quantum Hall system at $\nu=1/2$, the fermionic Moore--Read state is obtained from the following tunneling interaction Hamiltonian~\cite{teo2014luttinger,fuji2019quantum}
\be \label{eq:Pfaffian_Hamiltonian_1}
H=\sum_l\int dx\Big(\sum_{a,b=1}^2 t_{ab}\mathcal{O}^t_{l,ab}
+t_u\mathcal{O}_l^u+t_v\mathcal{O}_l^v\Big)+\text{H.c.},
\ee
where $\mathcal{O}^t_{l,ab}=\exp(i\phi^R_{l,a}-i\phi^L_{l+1,b})$ denotes the inter-pair tunneling operator, and $u,v$ denote intra-pair tunneling processes.
The fields entering these operators are constructed by grouping the microscopic wires into pairs labeled by $l$,
\be \label{eq:331_basis_1}
\phi^R_{l,1}&=&\varphi_{2l}+3\theta_{2l}+2\theta_{2l+1},
\nn
\phi^R_{l,2}&=&\varphi_{2l+1}+3\theta_{2l+1},
\nn
\phi^L_{l,1}&=&\varphi_{2l}-3\theta_{2l},
\nn
\phi^L_{l,2}&=&\varphi_{2l+1}-3\theta_{2l+1}-2\theta_{2l}.
\ee
We refer to these chiral variables as the $331$ basis.
They obey the commutation relations 
\be \label{eq:331_algebra_1}
[\phi^p_{l,a}(x),\phi^{p'}_{l',b}(x')]=i\pi p \delta_{pp'}\delta_{ll'}K_{ab}\sgn(x-x')
\ee
with $K$-matrix $K_{331}=((3,1),(1,3))$.
The intra-pair tunneling operators $\mathcal O_l^u$ and $\mathcal O_l^v$ are naturally expressed in terms of the charge and neutral combinations 
\be
\phi^c_{R/L,l}&=&\frac{1}{2}(\phi^{R/L}_{l,1}+\phi^{R/L}_{l,2}),
\nn
\phi^{n}_{R/L,l}&=&\frac{1}{2}(\phi^{R/L}_{l,1}-\phi^{R/L}_{l,2}),
\ee
for which $\mathcal O_l^{u/v} = \exp( i\phi^n_{R,l} \pm i\phi^n_{L,l})$~\cite{teo2014luttinger}.
To make the neutral sector gapping structure explicit, it is useful to refermionize the neutral field and decompose the resulting neutral Dirac fermion into two Majorana components~\cite{fuji2019quantum,teo2014luttinger},
\be
\psi^n_{R/L,l}&\equiv&\frac{\eta_{l}}{\sqrt{2\pi\alpha}}e^{i\phi^n_{R/L,l}}=\frac{1}{\sqrt{2}}(\xi^1_{R/L,l}+i\xi^2_{R/L,l}).
\ee
Following Ref.~\cite{teo2014luttinger}, we take $t_{ab}=t$, in which case
\be 
H&=&\sum_{l}\int dx \Big(4t \cos (\phi^c_{R,l}-\phi^c_{L,l+1})\xi^1_{R,l}\xi^1_{L,l+1}
\nn
&&+\frac{t_u-t_v}{2}i\xi_{R,l}^1\xi_{L,l}^1+\frac{t_u+t_v}{2}i\xi^2_{R,l}\xi^2_{L,l} \Big). 
\ee
This form separates the neutral sector into two Majorana channels with distinct intra-pair mass terms, while only one of them participates in the inter-pair coupling together with the charge sector. 
In the corresponding Pfaffian regime, the bulk is fully gapped, leaving at an open boundary one chiral charge mode and one co-propagating chiral Majorana mode~\cite{teo2014luttinger}.

In the continuum quantum Hall system, the tunneling operators entering the tunneling Hamiltonian Eq.~(\ref{eq:Pfaffian_Hamiltonian_1}) cannot all be made simultaneously non-oscillatory in a uniform magnetic field, and this construction therefore employs a spatially modulated magnetic field~\cite{teo2014luttinger}.
Here we reconsider the Pfaffian coupled-wire construction in the lattice setting, where crystal momentum is conserved modulo a longitudinal reciprocal lattice vector and can therefore accommodate tunneling processes that are momentum mismatched in the continuum.

{\it Lattice coupled-wire construction}---We begin with the direct lattice embedding of the $331$ basis [Eq.~(\ref{eq:331_basis_1})].
Setting $b=0$, the two pairs of chiral fields carry 
\be
(q^R_1,q^L_1)=(5k_F,-3k_F),
\quad
(q^R_2,q^L_2)=(3k_F,-5k_F).
\ee
Thus the momentum of $a=1,2$ fields differ by $2k_F$ in both chiral sectors. 
Consequently, the neutral fields $\phi^n_{R/L,l}$ carry finite crystal momentum $k_F$, and the neutral sector operators $\mathcal{O}^{u/v}_l$ are not all compatible with lattice momentum conservation in this basis.
The lattice setting, however, allows a different organization of the microscopic wire degrees of freedom. 
Because crystal momentum is conserved only modulo $2\pi/\lambda$ and is not tied to transverse wire separation through a magnetic-field-induced shift, tunneling processes between more distant wires remain allowed provided their net momentum transfer is commensurate with the reciprocal lattice. 
We therefore enlarge each two wire unit cell by inserting an auxiliary wire and use this additional degree of freedom to realign the momentum assignments of the two chiral field pairs.

\begin{figure}[htbp]
  \centering
  \resizebox{1.0\linewidth}{!}{
  %\documentclass{article}
%\usepackage{tikz}
%\usetikzlibrary{decorations.pathreplacing}
%\begin{document}
\begin{tikzpicture}
\draw[-,gray,thick](-3.5,-3.) -- (-2,-3.);
\draw[-,,gray,thick](-3.5,-1.5) -- (-2,-1.5);
\draw[-,gray,thick](-3.5,0.) -- (-2,0.);

\node at (-4.,-3.) {$\phi^{R/L}_{3l}$};
\node at (-4.,-1.5) {$\phi^{R/L}_{3l+1}$};
\node at (-4.,0.) {$\phi^{R/L}_{3l+2}$};

\draw[decorate, decoration={brace, amplitude=5pt}, thick] (-1.8,-1.5) -- (-1.8,-3.);

\draw[->,purple,thick] (-1.5,-2.25) -- (-0.5,-2.25);

\node at (0.25,-3.) {$\tilde{\phi}^{R/L}_{3l}$};
\node at (0.25,-1.5) {$\chi^{R/L}_{3l+1}$};
\node at (0.25,0.) {$\phi^{R/L}_{3l+2}$};

\draw[-,gray,thick](0.75,-3.) -- (2.25,-3.);
\draw[-,,gray,thick](0.75,-1.5) -- (2.25,-1.5);
\draw[-,gray,thick](0.75,0.) -- (2.25,0.);

\draw[decorate, decoration={brace, amplitude=5pt}, thick] (-0.2,-3) -- (-0.2,-1.5);

\node at (5.5,-3.) {${\phi}^{R/L}_{l,1}$};
\node at (5.5,-1.5) {$\chi^{R/L}_{3l+1}$};
\node at (5.5,0.) {${\phi}^{R/L}_{l,2}$};

\draw[-,gray,thick](3.5,-3.) -- (5,-3.);
\draw[-,,gray,thick](3.5,-1.5) -- (5,-1.5);
\draw[-,gray,thick](3.5,0.) -- (5,0.);

\draw[->, purple,thick] (2.5,-3) -- (3.25,-3);
\draw[->, purple,thick] (2.5,0) -- (3.25,0);
\draw[->, purple,thick] (2.5,-3) -- (3.25,-0.1);
\draw[->, purple,thick] (2.5,0) -- (3.25,-2.9);

%\draw[->,blue] (3,2) arc (350:0:0.1 and 0.2);
%\draw[gray, thick] (-1,2) -- (2,-4);
%\draw[gray, thick] (-1,-1) -- (2,2);
%\filldraw[black] (0,0) circle (2pt) node[anchor=west]{Intersection point};
\end{tikzpicture}
%\end{document}
}
  \caption{Schematic illustration of variables transformation.}
  \label{fig:1}
\end{figure}
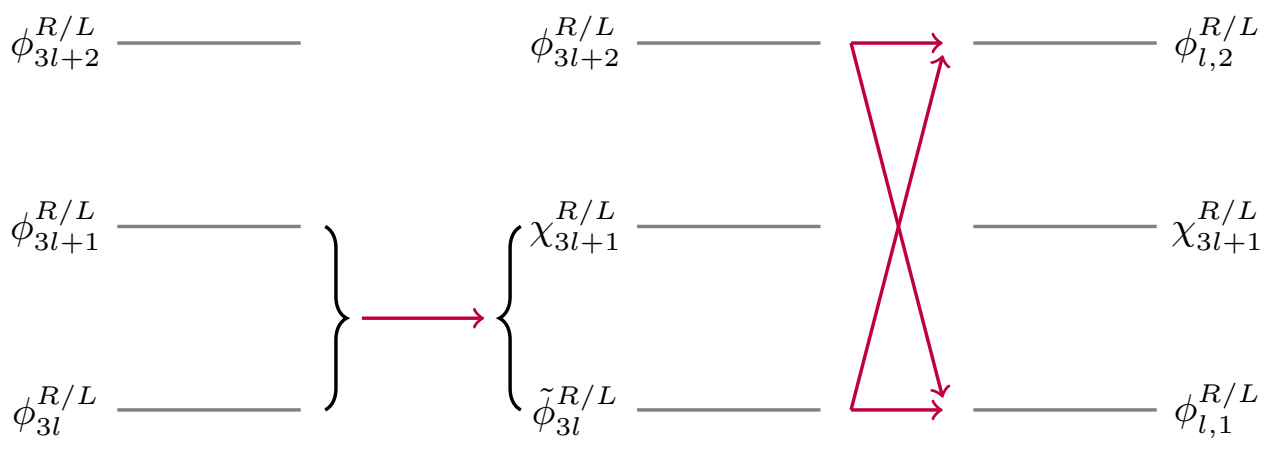

As illustrated in Fig.~\ref{fig:1}, we group the microscopic wires into three-wire unit cells. 
The auxiliary wire is introduced to dress one of the original microscopic channels so as to modify its crystal momentum assignment while preserving its chiral algebra. 
Within each unit cell, we perform the local transformation
\be
\tilde\phi^R_{3l}
&=&\varphi_{3l}+\theta_{3l}-2\theta_{3l+1}
\nn
\tilde\phi^L_{3l}
&=&\varphi_{3l}-\theta_{3l}-2\theta_{3l+1}
\nn
\chi^R_{3l+1}
&=&\varphi_{3l+1}+\theta_{3l+1}+2\theta_{3l}
\nn
\chi^L_{3l+1}
&=&\varphi_{3l+1}-\theta_{3l+1}+2\theta_{3l}.
\ee
The fields $\tilde\phi^{R/L}_{3l}$ describe the first microscopic electron channel dressed by a neutral particle-hole composite of the auxiliary wire, while $\chi^{R/L}_{3l+1}$ forms the complementary chiral channel.

We then apply the same $331$ transformation used in Ref.~\cite{teo2014luttinger} to the dressed wire $\tilde\phi^{R/L}_{3l}$ and the third microscopic wire $3l+2$. 
This yields
\be
\phi^R_{l,1}&=&\varphi_{3l}+3\theta_{3l}+2\theta_{3l+2}-2\theta_{3l+1}
\nn
\phi^R_{l,2}&=&\varphi_{3l+2}+3\theta_{3l+2}
\nn
\phi^L_{l,1}&=&\varphi_{3l}-3\theta_{3l}-2\theta_{3l+1}
\nn
\phi^L_{l,2}&=&\varphi_{3l+2}-3\theta_{3l+2}-2\theta_{3l}
\ee
together with an additional chiral pair,
\be
\phi^R_{l,3}&\equiv&\chi_{3l+1}^R=\varphi_{3l+1}+\theta_{3l+1}+2\theta_{3l}
\nn
\phi^L_{l,3}&\equiv& \chi_{3l+1}^L=\varphi_{3l+1}-\theta_{3l+1}+2\theta_{3l} 
\ee
The resulting fields obey the usual commutation relations with following $K$-matrix
\be
K_{\rm enlarged}
=
\begin{pmatrix}
3&1&0\\
1&3&0\\
0&0&1
\end{pmatrix}
=
K_{331}\oplus 1.
\ee
Thus the $a=1,2$ sector retains the $331$ chiral algebra, while the $a=3$ sector constitutes an additional algebraically decoupled $K=1$ charge channel.

We first consider the $a=1,2$ sector of the enlarged basis. 
Its key consequence is that the crystal momenta of the two $331$ components are aligned, $q^R_1=q^R_2=3k_F$ and $q^L_1=q^L_2=-5k_F$, so that the neutral fields $\phi^{R/L}_{l,1/2}$ carry vanishing crystal momentum.
This removes the momentum distinction among the different tunneling operators entering Eq.~(\ref{eq:Pfaffian_Hamiltonian_1}).
Writing 
\be
\phi^{R/L}_{l,a} = \phi^c_{R/L,l} +s_a\phi^n_{R/L,l},
\quad s_1=1, \ s_2=-1,
\ee
the inter-cell operators take the unified form
\be
\mathcal{O}^t_{l,ab}&=&\exp(i\phi^R_{l,a}-i\phi^L_{l+1,b})
\nn
&=&\exp (i(\phi^c_{R,l}-\phi^c_{L,l+1} +s_a\phi^n_{R,l} -s_b\phi^n_{L,l+1})).
\ee
Since the neutral fields carry no crystal momentum, all four operators $\mathcal O^t_{l,ab}$, $a,b=1,2$, have the same momentum transfer $8k_F$, which at half filling is an integer multiple of the reciprocal lattice vector $2\pi/\lambda$ and can therefore be absorbed by an Umklapp process.
The intra-cell operators $\mathcal O_l^{u/v}$ are likewise momentum conserving because they involve only the neutral fields.
Thus, the complete set of operators entering Eq.~(\ref{eq:Pfaffian_Hamiltonian_1}) is simultaneously allowed by lattice momentum conservation. 
It makes the full Pfaffian gapping structure accessible within the $a=1,2$ sector.

The remaining $a=3$ sector is an additional $K=1$ charge channel. 
Since the enlarged chiral algebra is block diagonal, $K_{\rm enlarged}=K_{331}\oplus 1$, this auxiliary sector can be gapped independently without modifying the $331$ sector.
One allowed interaction is the charge-density-wave (CDW) operator~\cite{kane2002fractional},
\be
\mathcal O_l^{\rm cdw}
&=&
\exp[
i(\phi^R_{l,3}-\phi^L_{l,3})
-i(\phi^R_{l+1,3}-\phi^L_{l+1,3})]
\nn
&=&
\exp[i(2\theta_{3l+1}-2\theta_{3l+4})].
\ee
This operator is local, charge conserving, and compatible with lattice momentum conservation, and hence is allowed in the microscopic interacting Hamiltonian~\cite{kane2002fractional,teo2014luttinger,fuji2019quantum}. 
Together with appropriate symmetry-allowed pinning interactions~\cite{shavit2024quantum}, the CDW coupling provides a possible route to gapping the auxiliary sector without introducing additional topologically protected boundary modes.
Importantly, the Pfaffian tunneling operators in the $a=1,2$ sector already satisfy crystal-momentum conservation modulo the reciprocal lattice vector of the original lattice, independently of whether the auxiliary sector develops CDW order.

With the $a=1,2$ sector realizing the Pfaffian gapping structure and the $a=3$ auxiliary sector independently gapped, the resulting boundary theory contains only one chiral charge mode and one co-propagating chiral Majorana mode.
At the level of momentum matching, the enlarged-wire construction thus plays the role served by the spatially modulated magnetic field in the continuum quantum Hall system~\cite{teo2014luttinger}. 
An analogous construction yields the anti-Pfaffian edge in the lattice setting which we elaborate in the supplements~\cite{supplemental_material}. 
Having established lattice realizations of both Pfaffian and anti-Pfaffian edges, we now turn to the Pf--aPf interface, where their microscopic momentum assignments constrain the allowed couplings between the edge degrees of freedom.

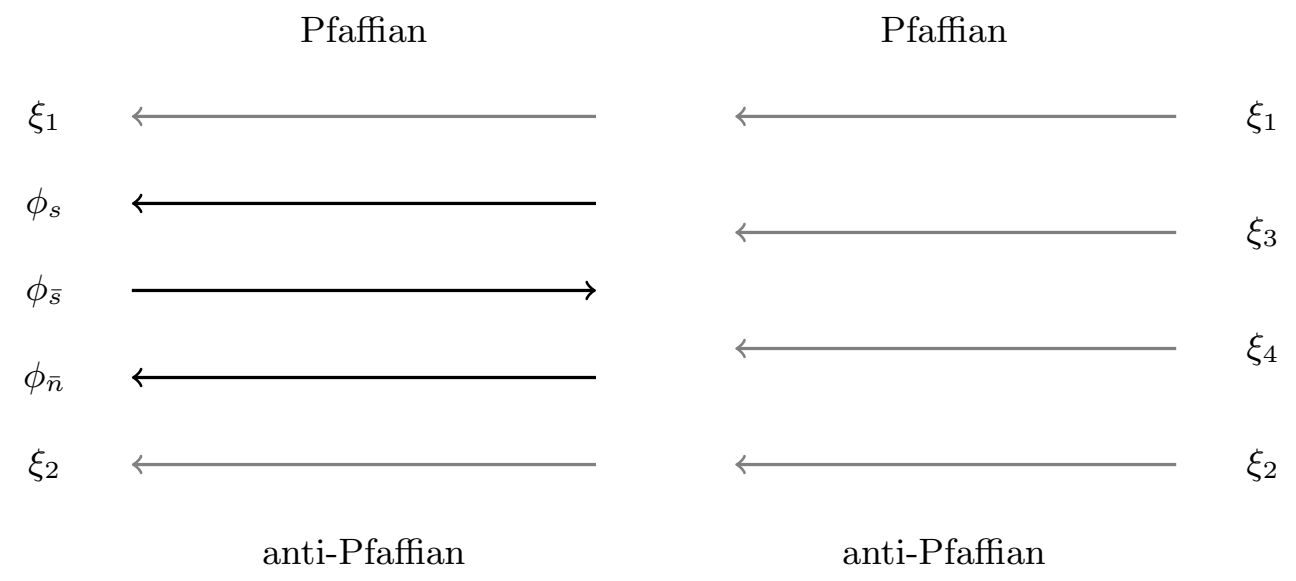
\begin{figure}[htbp]
  \centering
  \resizebox{1.0\linewidth}{!}{
  %\documentclass{article}
%\usepackage{tikz}
%\usetikzlibrary{decorations.pathreplacing}
%\begin{document}
\begin{tikzpicture}

\draw[<-,gray,thick](-4,2.25) -- (0,2.25);
\node at (-4.75,2.25) {$\xi_1$};

\draw[<-,black,thick](-4,1.5) -- (0,1.5);
\node at (-4.75,1.5) {$\phi_{s}$};

\draw[->,black,thick](-4,0.75) -- (0,0.75);
\node at (-4.75,0.75) {$\phi_{\bar{s}}$};

\node at (-4.75,0) {$\phi_{\bar{n}}$};
\draw[<-,black,thick](-4,0) -- (0,0);

\node at (-4.75,-0.75) {$\xi_2$};
\draw[<-,gray,thick](-4,-0.75) -- (0,-0.75);

\node at (-2,3.) {Pfaffian};
\node at (-2,-1.5) {anti-Pfaffian};

\draw[<-,gray,thick](1.2,2.25) -- (5,2.25);
\node at (5.75,2.25) {$\xi_1$};

\node at (5.75,1.25) {$\xi_3$};
\draw[<-,gray,thick](1.2,1.25) -- (5,1.25);

\node at (5.75,0.25) {$\xi_4$};
\draw[<-,gray,thick](1.2,0.25) -- (5,0.25);

\node at (5.75,-0.75) {$\xi_2$};
\draw[<-,gray,thick](1.2,-0.75) -- (5,-0.75);

\node at (3,3.) {Pfaffian};
\node at (3,-1.5) {anti-Pfaffian};

%\draw[decorate, decoration={brace, amplitude=5pt}, thick] (1.05,0.25) -- (1.05,1.25);

%\draw[->,red,thick](0.1,0) -- (0.8,0.75);

%\draw[->,blue] (3,2) arc (350:0:0.1 and 0.2);
%\draw[gray, thick] (-1,2) -- (2,-4);
%\draw[gray, thick] (-1,-1) -- (2,2);
%\filldraw[black] (0,0) circle (2pt) node[anchor=west]{Intersection point};
\end{tikzpicture}
%\end{document}
}
  \caption{Edge modes on the domain wall between Pf state and aPf state.}
  \label{fig:2}
\end{figure}

{\it Pfaffian--anti-Pfaffian domain wall}---
We now consider an interface between a Pf domain and an aPf domain, as illustrated in Fig.~\ref{fig:2}. 
As shown there, the low energy interface theory contains two counter-propagating charged chiral bosons, $\phi_s$ and $\phi_{\bar s}$, a neutral chiral boson $\phi_{\bar{n}}$ from the aPf edge, and two neutral chiral Majorana fermions, $\xi_1$ and $\xi_2$, originating from the Pf and aPf sectors, respectively. 
The interface fields are related to the microscopic coupled-wire variables using the conventional quasiparticle normalization~\cite{teo2014luttinger}, as detailed in the Supplemental Material~\cite{supplemental_material}.

In this lattice coupled-wire construction, the counter-propagating charge modes can be fully gapped by a local, charge conserving interaction. 
A suitable interaction Hamiltonian is~\cite{lian2018theory,barkeshli2015particle}
\be
H_{\rm int}=\int dx g \exp(4i(\phi_s+\phi_{\bar{s}}))+\text{H.c.}
\ee
where $g$ is corresponding coupling constant.

In terms of the microscopic wire degrees of freedom, this operator is a local product of electron operators and carries momentum $16k_F$. 
The operator is therefore momentum conserving modulo a reciprocal lattice vector at $\nu=1/2$. 
Its scaling dimension depends on the forward scattering interactions in the charge sector and can be made relevant in an appropriate interaction regime~\cite{lian2018theory}. 
When this term flows to strong coupling, the non-chiral charge modes $\phi_s$, $\phi_{\bar{s}}$ are gapped and the interface becomes purely neutral at low energies. 
The lattice coupled-wire construction thus provides an explicit microscopic realization of the neutral Pf--aPf domain wall building block used in thermal transport and network descriptions~\cite{wan2016striped,wang2018topological,mross2018theory,lian2018theory}.

After the charge sector is gapped, the remaining neutral bosonic mode of the aPf edge can be refermionized into two chiral Majorana fermions, $\xi_3$ and $\xi_4$~\cite{mross2018theory,lian2018theory,lee2007particle,levin2007particle}.
Together with the Majorana modes $\xi_1$ and $\xi_2$ inherited from the Pf and aPf sectors, the low energy domain wall theory therefore contains four co-propagating chiral Majorana modes, which are further distinguished by their microscopic momentum structure. 
The Majorana modes $\xi_1$ and $\xi_2$ originate from neutral sectors centered at zero momentum, whereas the neutral Dirac mode that gives rise to $\xi_3$ and $\xi_4$ is centered at momentum 
$-k_F$~\cite{supplemental_material}.
The two sets of Majorana fields therefore transform under distinct representations of lattice translation symmetry, which forbids translation-invariant bilinear mixing between them at a clean interface. 
Equivalently, such mixing requires a finite momentum transfer of order $k_F$~\cite{wang2018topological}. 
By contrast, bilinear mixing within the pairs $(\xi_1,\xi_2)$ and $(\xi_3,\xi_4)$ is compatible with translation symmetry.
The four Majorana modes thus naturally organize into a $2+2$ block structure for this microscopic realization~\footnote{This separation persists even if the auxiliary sector develops a period-doubled CDW. The two Majorana pairs remain separated by $k_F$, which is nonzero modulo the reciprocal lattice vector $2k_F$ of the doubled unit cell, so their bilinear mixing remains forbidden by the residual translation symmetry.}.

The resulting $2+2$ structure provides a microscopic basis for the pairwise mixing structure considered in effective theories of Pfaffian--anti-Pfaffian domain walls, whose quadratic dynamics can be described by~\cite{lian2018theory,wang2018topological,mross2018theory}
\be \label{eq:majorana_theory}
\mathcal{L}=\sum_{I=1}^{4} \xi_{I}(i\partial_t-iv_I\partial_x)\xi_I.
\ee
If the velocities are degenerate within each momentum sector, $v_1=v_2$ and $v_3=v_4$, the low energy Majorana theory Eq.~(\ref{eq:majorana_theory}) is invariant under independent $O(2)$ rotations of $(\xi_1,\xi_2)$ and $(\xi_3,\xi_4)$, giving an $O(2)\times O(2)$ flavor symmetry~\cite{wang2018topological}.
The two velocity degeneracies, however, have different microscopic origins. 
The equality $v_3=v_4$ arises naturally because $\xi_3$ and $\xi_4$ are the two Majorana components of the same neutral Dirac mode~\cite{lian2018theory}. 
By contrast, no analogous microscopic constraint enforces $v_1=v_2$, since $\xi_1$ and $\xi_2$ originate from the Pf and aPf sides of the interface, respectively. 
A finite velocity anisotropy $\delta v=v_1-v_2$ therefore breaks the exact $O(2)$ symmetry of the first pair~\cite{lian2018theory}.
Such a velocity anisotropy, however, does not by itself remove the momentum-space separation between the two Majorana pairs. 
For sufficiently weak anisotropy, the clean lattice realization therefore remains naturally organized around a $2+2$ Majorana structure, even though the exact $O(2)\times O(2)$ flavor symmetry is absent~\footnote{Additional symmetry-allowed couplings, including intra-pair Majorana bilinears and quartic interactions, can modify the low-energy dynamics and the exact flavor symmetry, but do not alter the momentum-selection rule for inter-pair bilinear mixing.}.

Disorder provides a natural mechanism for modifying both the momentum selection rules and the velocity anisotropies of the Majorana theory, as discussed previously in studies of Pf--aPf domain walls~\cite{lian2018theory,mross2018theory,wang2018topological}. 
On the one hand, velocity anisotropies within a Majorana pair can become irrelevant near disorder-dominated fixed points~\cite{lian2018theory}, promoting an approximate $O(2)\times O(2)$ flavor symmetry. 
On the other hand, introducing disorder permits scattering between the two momentum sectors~\cite{kane1994randomness,wang2018topological,lian2018theory}, thereby weakening the $2+2$ separation and opening the possibility of a more strongly mixed, potentially $O(4)$ flavor symmetry.
Determining which flavor symmetry regime emerges from the present microscopic realization, and how it shapes the corresponding disordered phase diagram, requires a more detailed analysis of the disorder-driven crossover, which we leave for future work.

{\it Conclusion}---In summary, we developed lattice coupled-wire constructions of the Pfaffian and anti-Pfaffian states in half-filled moir\'{e} systems without relying on the nonuniform magnetic-field or wire-spacing arrangement used to satisfy momentum matching in the corresponding continuum construction.
The construction retains the microscopic momentum structure of the boundary degrees of freedom, which we used to analyze a Pfaffian--anti-Pfaffian domain wall, including the momentum-allowed gapping of its charge sector and the organization of its neutral Majorana modes. 
Our results show that lattice crystal momentum conservation permits coupled-wire realizations beyond those accessible in the corresponding uniform-field, equally spaced-wire continuum setting.
This suggests that lattice momentum constraints provide an additional organizing principle for non-Abelian FQAH phases and may enable analogous constructions beyond the Pfaffian family~\cite{fuji2017non,tam2020coupled,kane2018coupled}.

\hfill

{\it Acknowledgments}--- We acknowledge the support of ECS project 26309524 of the research grant council of Hong Kong.

\bibliography{E1}

\onecolumngrid

\onecolumngrid

\newpage
\clearpage

% \allowdisplaybreaks
\vspace{1cm}
\begin{center}
\textbf{\large Supplemental material}
\end{center}

\setcounter{secnumdepth}{3}
\setlength\parindent{0pt}

\setcounter{equation}{0}
\setcounter{figure}{0}
\setcounter{table}{0}
\setcounter{page}{1}

\makeatletter
\renewcommand{\theequation}{S\arabic{equation}}
\renewcommand{\thefigure}{S\arabic{figure}}
\renewcommand{\thetable}{S\arabic{table}}
\renewcommand{\thesection}{S\arabic{section}}
\newcommand{\lpartial}{\overleftarrow{\partial}}
\newcommand{\rpartial}{\overrightarrow{\partial}}
\newcommand{\lnabla}{\overleftarrow{\nabla}}
\newcommand{\rnabla}{\overrightarrow{\nabla}}

\section{Coupled-wires construction of anti-Pfaffian state}

Compared with the Pfaffian state, the anti-Pfaffian state supports an additional neutral edge mode, suggesting that three wires should be grouped together in its coupled-wire construction. 
More explicitly, for each three-wire unit we define the $113$ variables
\be
\phi^R_{l,1}&=&\varphi_{3l}+\theta_{3l}+2\theta_{3l+1}+6\theta_{3l+2}
\nn
\phi^R_{l,2}&=& \varphi_{3l+2}+\theta_{3l+2}+2\theta_{3l+1}
\nn
\phi^L_{l,1}&=&\varphi_{3l}-\theta_{3l}-2\theta_{3l+1}
\nn
\phi^L_{l,2}&=&\varphi_{3l+2}-\theta_{3l+2}-2\theta_{3l+1}-6\theta_{3l}
\ee
together with a pair of neutral variables
\be \label{eq:neutral_variables}
\tilde{\phi}^{n}_{R,l}&=&\frac{1}{2}\Big(-\varphi_{3l}+2\varphi_{3l+1}-\varphi_{3l+2}+3\theta_{3l}-2\theta_{3l+1}-3\theta_{3l+2}
\Big)
\nn
\tilde{\phi}^{n}_{L,l}&=&\frac{1}{2}\Big(\varphi_{3l}-2\varphi_{3l+1}+\varphi_{3l+2}-3\theta_{3l}-2\theta_{3l+1}+3\theta_{3l+2}
\Big)
\ee
These two sectors mutually commute and satisfy the commutation relations
\be
&&\text{[}\phi^R_{l,a}(x),\phi^R_{l,b}(x')]=i\pi \delta_{l,l'}K_{ab}\sgn(x-x'), \qquad
\text{[}\phi^L_{l,a}(x),\phi^L_{l,b}(x')]=-i\pi \delta_{l,l'}K_{ab}\sgn(x-x'),
\nn
&&\text{[}\tilde{\phi}^{n}_{R,l}(x),\tilde{\phi}^{n}_{R,l'}(x')]=-i\pi \delta_{l,l'}\sgn(x-x'),
\qquad
\text{[}\tilde{\phi}^{n}_{L,l}(x),\tilde{\phi}^{n}_{L,l'}(x')]=i\pi \delta_{l,l'}\sgn(x-x'),
\nn
&&\text{[}\phi^R_{l,a}(x),\phi^L_{l,b}(x')]=0,
\qquad \qquad \qquad \qquad \qquad \ 
\text{[}\tilde{\phi}^{n}_{R,l}(x),\tilde{\phi}^{n}_{L,l'}(x')]=0,
\ee
with $K$-matrix
\be
K=\left(
\begin{matrix}
    1&3\\3&1
\end{matrix}
\right)
\ee
In terms of these bosonic field, the tunneling Hamiltonian can be written as 
\be \label{eq:anti_Pfaffian_Hamiltonian_1}
H&=&\int dx \sum_l \Big(4t\exp(i\phi^c_{R,l}-i\phi^c_{L,l+1})\cos(\phi^n_{R,l})\cos(\phi^n_{L,l+1})+t_u\exp(i({\phi}^n_{R,l}+{\phi}^n_{L,l})
)+t_v\exp(i({\phi}^n_{R,l}-{\phi}^n_{L,l}))
\nn
&&+2\tilde{t}\exp(i\phi^c_{R,l}-i\phi^c_{L,l+1})\cos (\tilde{\phi}^n_{R,l}-\tilde{\phi}^n_{L,l+1})+\text{H.c.}\Big)
\ee
with
\be
\phi^c_{R/L,l}=\frac{1}{2}(\phi^{R/L}_{l,1}+\phi^{R/L}_{l,2}),
\qquad
\phi^n_{R/L,l}=\frac{1}{2}(\phi^{R/L}_{l,1}-\phi^{R/L}_{l,2}).
\ee
This gapping pattern yields the anti-Pfaffian phase, with a fully gapped bulk and an open boundary supporting one chiral charge mode, one counter-propagating neutral mode, and one counter-propagating chiral Majorana mode~\cite{teo2014luttinger,fuji2019quantum}.
However, similar to Pfaffian case, operators in the tunneling Hamiltonian Eq.~(\ref{eq:anti_Pfaffian_Hamiltonian_1}) cannot all be made simultaneously non-oscillatory in a uniform magnetic field.
A spatially modulated magnetic field is therefore required to achieve momentum matching for all tunneling processes.

In a lattice system, by adding an extra wire between $3l$ and $3l+1$, we enlarge each three-wire unit cell into four-wire unit cell and dress $\phi^{R/L}_{l,1}$ channel so as to modify its crystal momentum assignment without changing the underlying chiral commutation algebra
\be
\phi^R_{l,1}&=&\varphi_{4l}+\theta_{4l}+2\theta_{4l+2}+6\theta_{4l+3}-6\theta_{4l+1}
\nn
\phi^L_{l,1}&=&\varphi_{4l}-\theta_{4l}-2\theta_{4l+2}-6\theta_{4l+1}
\nn
\phi^R_{l,2}&=& \varphi_{4l+3}+\theta_{4l+3}+2\theta_{4l+2}
\nn
\phi^L_{l,2}&=&\varphi_{4l+3}-\theta_{4l+3}-2\theta_{4l+2}-6\theta_{4l}
\ee
together with an additional chiral pair
\be
\phi^R_{l,3}&=&\varphi_{4l+1}+\theta_{4l+1}+6\theta_{4l}+3\theta_{4l+2}
\nn
\phi^L_{l,3}&=&\varphi_{4l+1}-\theta_{4l+1}+6\theta_{4l}+3\theta_{4l+2}
\ee
These variable commute with neutral variables [Eq.~(\ref{eq:neutral_variables})] and satisfy usual commutation relations with following $K$-matrix
\be
K=\left(\begin{matrix}
    1&3&0\\3&1&0\\0&0&1
\end{matrix}
\right)
\ee
The tunneling Hamiltonian can be expressed as 
\be \label{eq:anti_Pfaffian_Hamiltonian}
H&=&\int dx \sum_l \Big(4t\exp(i\phi^c_{R,l}-i\phi^c_{L,l+1})\cos(\phi^n_{R,l})\cos(\phi^n_{L,l+1})+t_u\exp(i({\phi}^n_{R,l}+{\phi}^n_{L,l})
)+t_v\exp(i({\phi}^n_{R,l}-{\phi}^n_{L,l})
)
\nn
&&+4\tilde{t}\exp(i\phi^c_{R,l}-i\phi^c_{L,l+1})\cos (\tilde{\phi}^n_{R,l}-\tilde{\phi}^n_{L,l+1})
\cos\Big(\frac{1}{2}((\phi^R_{l,3}-\phi^L_{l,3})-(\phi^{R}_{l+1,3}-\phi^L_{l+1,3}))\Big)+\text{H.c.}\Big)
\ee
Analogous to the Pfaffian case, the complete set of operators appearing in Eq.~(\ref{eq:anti_Pfaffian_Hamiltonian}) is local, charge conserving, and simultaneously allowed by lattice momentum conservation. The resulting boundary theory contains one chiral charge mode, one counter-propagating neutral mode, and one counter-propagating chiral Majorana mode~\cite{teo2014luttinger,fuji2019quantum}.
In particular, when the $a=1,2$ and neutral variable $\tilde{\phi}^n$ sectors are integrated out, we will have a residual tunneling term consisting of auxiliary sector~\cite{fuji2019quantum}
\be
\cos\Big((\phi^R_{l,3}-\phi^L_{l,3})-(\phi^{R}_{l+1,3}-\phi^L_{l+1,3})\Big).
\ee
This tunneling term produces the charge density wave state~\cite{kane2002fractional} and does not introduce additional protected boundary modes.

\section{Microscopic Structure of the Pfaffian–Anti-Pfaffian Interface}

We consider the interface geometry shown in Fig.~2, with the Pfaffian region above the anti-Pfaffian region.
We denote the reference wire indices of the boundary on the Pfaffian and anti-Pfaffian sides by $j_{\mathrm{Pf}}$ and $j_{\mathrm{aPf}}$, respectively.

We first identify the microscopic edge fields on the Pfaffian side. 
At the lower boundary of the Pfaffian region, the two surviving left-moving $331$ fields are denoted by $\phi_a\equiv\phi^L_{l,a}$, $a=1,2$, with $3l=j_{\mathrm{Pf}}$. 
They obey
\be
[\phi_a(x),\phi_b(x')]
=-i\pi(K_{331})_{ab}\operatorname{sgn}(x-x'),
\qquad
K_{331}=
\begin{pmatrix}
3&1\\
1&3
\end{pmatrix}.
\ee
Following the quasiparticle normalization of the reference construction~\cite{teo2014luttinger}, we introduce
\be
\Phi_a=\sum_{b=1}^{2}(K_{331}^{-1})_{ab}\phi_b,
\qquad
K_{331}^{-1}=\frac{1}{8}
\begin{pmatrix}
3&-1\\
-1&3
\end{pmatrix}.
\ee
The corresponding charge and neutral combinations are
\be
\phi_s=\Phi_1+\Phi_2=\frac14(\phi_1+\phi_2),
\qquad
\phi_n=\Phi_1-\Phi_2=\frac12(\phi_1-\phi_2).
\ee
Here $\exp(i\phi_s)$ represents a charge-$e/2$ Abelian quasiparticle, while $\exp(i\phi_n)$ is electrically neutral. 
The neutral-sector gapping in the Pfaffian construction leaves the charge sector unchanged and retains a single chiral Majorana component of the refermionized neutral field.
Substituting the coupled-wire expressions for $\phi^L_{l,1}$ and $\phi^L_{l,2}$, we obtain
\be
\phi_{s}&=&\frac{1}{4}\Big(\varphi_{j_{\mathrm{Pf}}}+\varphi_{j_{\mathrm{Pf}}+2}-5\theta_{j_{\mathrm{Pf}}}-3\theta_{j_{\mathrm{Pf}}+2}- 2\theta_{j_{\mathrm{Pf}}+1}\Big)
\nn
\phi_n&=&\frac{1}{2}\Big(\varphi_{j_{\mathrm{Pf}}}-\varphi_{j_{\mathrm{Pf}}+2}-\theta_{j_{\mathrm{Pf}}}+3\theta_{j_{\mathrm{Pf}}+2}- 2\theta_{j_{\mathrm{Pf}}+1}
\Big)
\ee
These fields satisfy
\be
&&[\phi_{s}(x),\phi_s(x')]=-i\pi\frac{1}{2}\sgn(x-x')
\nn
&&\text{[}\phi_n(x),\phi_n(x')]=-i\pi \sgn(x-x')
\ee
The charge and neutral sectors are therefore co-propagating and left-moving. 
The vertex operator $\exp(i\phi_s)$ carries crystal momentum $-5k_F/2$, while the neutral sector gives rise to a surviving Majorana mode $\xi_1$ centered at zero crystal momentum, as used in the main text.

\hfill

We next consider the upper boundary of the anti-Pfaffian region. Using the analogous quasiparticle normalization, with the charge-field convention chosen such that $\exp(-i\phi_{\bar s})$ represents a charge-(e/2) quasiparticle, we obtain the following microscopic boundary fields:
\be
\phi_{\bar{s}}&=&-\frac{1}{4}\Big(\varphi_{j_{\mathrm{aPf}}}+\theta_{j_{\mathrm{aPf}}}+\varphi_{j_{\mathrm{aPf}}+3}+7\theta_{j_{\mathrm{aPf}}+3}+4\theta_{j_{\mathrm{aPf}}+2}-6\theta_{j_{\mathrm{aPf}}+1}\Big)
\nn
\tilde{\phi}_{\bar{n}}&=&\frac{1}{2}\Big(\varphi_{j_{\mathrm{aPf}}}+\theta_{j_{\mathrm{aPf}}}-\varphi_{j_{\mathrm{aPf}}+3}+5\theta_{j_{\mathrm{aPf}}+3}-6\theta_{j_{\mathrm{aPf}}+1}
\Big)
\nn
\phi_{\bar{n}}&=&\frac{1}{2}\Big(-\varphi_{j_{\mathrm{aPf}}}+3\theta_{j_{\mathrm{aPf}}}+2\varphi_{j_{\mathrm{aPf}}+2}-2\theta_{j_{\mathrm{aPf}}+2}-\varphi_{j_{\mathrm{aPf}}+3}-3\theta_{j_{\mathrm{aPf}}+3}
\Big)
\ee
These fields obey
\be
&&[\phi_{\bar{s}}(x),\phi_{\bar{s}}(x')]=i\pi\frac{1}{2}\sgn(x-x')
\nn
&&\text{[}\tilde{\phi}_{\bar{n}}(x),\tilde{\phi}_{\bar{n}}(x')]=-i\pi\sgn(x-x')
\nn
&&\text{[}\phi_{\bar{n}}(x),\phi_{\bar{n}}(x')]=-i\pi\sgn(x-x')
\ee
with vanishing mutual commutators.
The charge field $\phi_{\bar s}$ represents the counter-propagating charge sector, with $e^{-i\phi_{\bar s}}$ corresponding to a charge $e/2$ Abelian quasiparticle carrying crystal momentum $3k_F/2$. 
The two neutral bosonic variables describe distinct sectors: $\tilde{\phi}_{\bar{n}}$ enters the neutral-sector gapping procedure that leaves a single chiral Majorana mode $\xi_2$, whereas $\phi_{\bar{n}}$ remains as a neutral chiral boson. 
Their microscopic momentum assignments are $0$ and $-k_F$, respectively. 
The commutation relations establish that both neutral sectors propagate oppositely to the charge sector~\cite{fuji2019quantum}.

\hfill

Having identified the microscopic boundary fields on both sides of the interface, we now connect them to the effective Pfaffian--anti-Pfaffian interface theory used in Ref.~\cite{lian2018theory}. 
The bosonic sectors can be expressed in terms of the fields
\be
\phi_0=2\phi_{\bar s}-\phi_{\bar n},
\qquad
\phi_1=\phi_s,
\qquad
\phi_2=\phi_{\bar n}-\phi_{\bar s}.
\ee
Their commutation relations correspond to the $K$ matrix
\be
K=\text{diag}(-1,2,2),
\ee
which gives the effective interface action $S=S_{\phi}+S_{\xi}$, 
\be
S_{\phi}&=&(4\pi)^{-1}\sum_{I,J=0}^2\int dtdx (K_{IJ}\partial_t\phi_I\partial_x\phi_J-V_{IJ}\partial_x\phi_I\partial_x\phi_J)
\nn
S_{\xi}&=&\sum_{I=1}^2\int dtdx \xi_I(i\partial_t-iv_{I}\partial_x)\xi_I
\ee
Here $V_{IJ}$ parametrizes the interactions and velocities in the bosonic sector, while $\xi_1$ and $\xi_2$ are the surviving neutral Majorana modes from the Pfaffian and anti-Pfaffian boundaries, respectively.

We now consider the charge-gapping interaction
\be
H_{\rm int}=\int dx g \exp(4i(\phi_s+\phi_{\bar{s}}))+\text{H.c.}
\ee
Using the microscopic expressions of the boundary fields, the interaction operator can be written as
\be
\exp(4i(\phi_s+\phi_{\bar{s}}))&=&\exp(i\Big(
\varphi_{j_{\mathrm{Pf}}}+\varphi_{j_{\mathrm{Pf}}+2}-5\theta_{j_{\mathrm{Pf}}}-3\theta_{j_{\mathrm{Pf}}+2}- 2\theta_{j_{\mathrm{Pf}}+1}
\Big)
\nn
&&-i\Big(\varphi_{j_{\mathrm{aPf}}}+\theta_{j_{\mathrm{aPf}}}+\varphi_{j_{\mathrm{aPf}}+3}+7\theta_{j_{\mathrm{aPf}}+3}+4\theta_{j_{\mathrm{aPf}}+2}-6\theta_{j_{\mathrm{aPf}}+1}\Big))
\nn
&\sim&\psi_{L,j_{\mathrm{Pf}}}\psi_{L,j_{\mathrm{Pf}}+2}(\psi_{L,j_{\mathrm{Pf}}}\psi^{\dagger}_{R,j_{\mathrm{Pf}}})^2(\psi_{L,j_{\mathrm{Pf}}+2}\psi^{\dagger}_{R,j_{\mathrm{Pf}}+2})(\psi_{L,j_{\mathrm{Pf}}+1}\psi^{\dagger}_{R,j_{\mathrm{Pf}}+1})
\nn
&&\otimes \Big(\psi_{R,j_{\mathrm{aPf}}}\psi_{R,j_{\mathrm{aPf}}+3}(\psi_{R,j_{\mathrm{aPf}}+3}\psi^{\dagger}_{L,j_{\mathrm{aPf}}+3})^3(\psi_{R,j_{\mathrm{aPf}}+2}\psi^{\dagger}_{L,j_{\mathrm{aPf}}+2})^2(\psi_{L,j_{\mathrm{aPf}}+1}\psi^{\dagger}_{R,j_{\mathrm{aPf}}+1})^3
\Big)^{\dagger}
\ee
where the explicit coefficients and electron operators are specified by the wire embedding. 
This expression establishes the electron locality of the interaction. 
Its net crystal-momentum transfer has magnitude $16k_F$, which at half filling is an integer multiple of the reciprocal lattice vector $2\pi/\lambda$. 
The interaction is therefore allowed by lattice translation symmetry.
The charge-gapping combination also satisfies the null-vector condition,
\be
[\phi_s(x)+\phi_{\bar s}(x),
\phi_s(x')+\phi_{\bar s}(x')]=0.
\ee
Its relevance can be analyzed using the effective interface theory of Ref.~\cite{lian2018theory}. 
For the interaction parameters satisfying $V_{11}=V_{22}$ and $V_{01}=V_{02}$, the scaling dimension of the interaction operator is
\be
\Delta_{\mathcal O}
=8\sqrt{\frac{2V_{00}+V_{11}+V_{12}-4V_{01}}
{2V_{00}+V_{11}+V_{12}+4V_{01}}}.
\ee
For $\Delta_{\mathcal O}<2$, the interaction is relevant and flows to strong coupling, gapping the counter-propagating charge modes. 
Suitable forward-scattering interactions can realize this regime~\cite{lian2018theory}.

\hfill

After the charge sector is gapped, the surviving neutral degrees of freedom are the Majorana modes $\xi_1,\xi_2$ and the chiral neutral boson $\phi_{\bar n}$.
Refermionizing the latter into a neutral Dirac field $\Psi_{\bar{n}}=(\xi_3+i\xi_4)/\sqrt{2}$, the microscopic momentum assignments show that $\xi_1,\xi_2$ transform trivially under lattice translation, whereas $\xi_3,\xi_4$ form a two-dimensional subspace rotated by an angle $\pi/2$ under translation by one lattice period. 
The four Majorana modes therefore divide into two distinct translation representations, giving the $2+2$ structure discussed in the main text.

\end{document}